\magnification=1200
\font\tenbf=cmbx10
\font\tenrm=cmr10
\font\tenit=cmti10

\font\eightbf=cmbx8
\font\eightrm=cmr8
\font\eightit=cmti8

\def\lsim{\raise0.3ex\hbox{$<$\kern-0.75em\raise-1.1ex\hbox{$\sim$}}}
\def\gsim{\raise0.3ex\hbox{$>$\kern-0.75em\raise-1.1ex\hbox{$\sim$}}}
\def\binum{\hbox{BI-TP 94/53 \strut}}
\def\datnum{\hbox{October 1994\strut}}
\def\banner{\hfill\hbox{\vbox{\offinterlineskip
                              \binum\datnum}}\relax}
\nopagenumbers

\hsize=6.0truein
\vsize=8.5truein
\parindent=15pt
\baselineskip=10pt
\def\qed{\hbox{${\vcenter{\vbox{
    \hrule height 0.4pt\hbox{\vrule width 0.4pt height 6pt
    \kern5pt\vrule width 0.4pt}\hrule height 0.4pt}}}$}}
\hfill\banner

\vglue 1pc
\baselineskip=13pt
\centerline{\tenbf HADRON PROPERTIES IN THE VICINITY OF $T_c$\footnote{
$^\dagger$}{\eightrm Invited talk given at the International Workshop on
Multi-Particle Correlations and Nuclear Reactions, {\eightit CORINNE II},
September 6-10 1994, Nantes, France.}}
\vglue 24pt
\centerline{\tenrm FRITHJOF KARSCH}
\baselineskip=12pt
\centerline{\eightit Fakult\"at f\"ur Physik, Universit\"at Bielefeld}
\baselineskip=10pt
\centerline{\eightit Universit\"atsstr., D-33615 Bielefeld, Germany}
\vglue 16pt
\centerline{\tenrm ABSTRACT}
{\rightskip=3.0pc
 \leftskip=3.0pc
 \eightrm\baselineskip=12pt\parindent=1pc
Modifications of hadron masses and some of their basic properties with
temperature or nuclear density are considered as one possible signatures
for the formation of dense hadronic matter in nuclear collisions. We
discuss here some basic results on the temperature dependence of hadron
properties obtained from calculations of hadron correlation functions, the
chiral condensate as well as the equation of state
in finite temperature lattice QCD.
\vglue 12pt}
\tenrm\baselineskip=14pt
\line{\tenbf 1. Introduction \hfil}
\vglue 5pt

The theory of strongly interacting particles -- Quantumchromodynamics
(QCD) -- has many unusual non-perturbative features, which are under
intensive theoretical and
experimental investigation. It is the hope that some of the low energy
characteristics of QCD like confinement and chiral symmetry breaking may
become better understood through studies of strongly interacting matter,
{\sl ie.} large systems of hadrons at high temperature and/or
high baryon number density. This is, of course, also of relevance for our
understanding of the evolution of the early universe.

Properties of strongly interacting matter are studied in
ultra-relativistic heavy ion experiments. Through an analysis of the
final state hadrons one tries to extract information about the dense
thermal medium, created in these collisions.
Basic properties of hadrons like their masses, widths and decay constants
reflect many features of the complex, non-perturbative structure of the
QCD vacuum. Any modifications of these hadronic properties at high
temperature and/or density may lead to modifications in the particle spectra,
which will lead to experimentally observable effects. It thus is important
to understand theoretically the equilibrium properties of hadrons in
dense/hot nuclear media, although the detailed description  of heavy 
ion
experiments will still involve further assumptions about non-equilibrium
effects and the thermal evolution in a heavy ion collision$^1$.

Effective theories, deduced from QCD in order to describe low energy
properties of hadrons, establish a close link between hadronic
properties and the non-perturbative structure of the QCD vacuum, which is
described by various non-vanishing condensates. This is, for instance,
used in the operator product expansion (OPE) for correlation functions
of hadronic currents$^2$, which allows to relate hadron masses to
condensates of the quark and gluon fields. The temperature and density
dependence of the latter has been discussed within the context of chiral
perturbation theory$^3$. Through an extension of the OPE to finite
temperature it is then possible to
discuss the temperature dependence of hadron masses
and other hadronic properties in a nuclear medium. We should, however,
mention that this approach is not at all straightforward and led to some
controversial discussion about the relevant states which should be used
in the finite temperature expansion$^4$.

We will concentrate here on a discussion of the influence of non-zero
temperatures on hadronic properties\footnote{$^\dagger$}{\eightrm 
For a discussion 
of the influence of a non-vanishing baryon density we refer to the 
contribution of W. Weise to these proceedings.}. 
The applicability of the OPE as well as chiral
perturbation theory is limited to low temperatures. In the vicinity of
the QCD phase transition from hadronic matter to the quark gluon plasma
phase, however, the hadronic medium becomes so dense that the influence
of many particle states and resonances can no
longer be neglected. In this regime an entirely non-perturbative approach
becomes mandatory. This is the regime where lattice QCD might provide
the only reliable approach to study the temperature dependence of
condensates, hadrons and their properties.

We will discuss here some of the recent results of lattice calculations,
which aim at an analysis of hadronic properties close to the QCD phase
transition. In the next section we briefly review results about the
thermodynamics of QCD, focusing on the equation of state and chiral
properties in the vicinity of $T_c$. Section 3 is devoted to a
discussion of lattice calculations of hadron masses and decay constants
at finite temperature within the quenched approximation. We briefly
comment about the situation in two-flavour QCD in Section 4.
Finally we give our conclusions in Section 5.
\vglue 12pt
\line{\tenbf 2. Basic Thermodynamics from Lattice QCD \hfil}
\vglue 5pt

Thanks to the rapid development of computer technology and the improvement
of numerical algorithms for Monte Carlo calculations the
studies of lattice QCD at finite temperature lead to a
steadily improving understanding of details of the QCD phase transition.
In the absence of dynamical quark degrees of freedom (pure gauge
theory/quenched QCD) we have obtained quantitative results for the order
of the transition, the transition temperature and equation of state on
lattices of different size. This allows with some confidence an
extrapolation to the continuum limit$^{5,6}$. Although simulations with
light quarks did not yet reach a similar accuracy we know also here a lot
about the qualitative behaviour of QCD at finite temperature.

As discussed in the introduction we expect that the properties of
hadrons at finite temperature are closely related to changes in
non-perturbative condensates in the QCD vacuum. With increasing temperature
more and more hadronic states get excited and populate the vacuum.
This is expected to result in decreasing values for
the condensates. Finally, when the density of hadrons becomes too large
the whole space is filled by hadronic bubbles which completely suppress the
non-perturbative chiral condensate --
the phase transition to the quark gluon plasma phase occurs and the chiral
symmetry is restored above $T_c$. 

This simplified picture of the QCD phase transition suggests a similarity
with a percolation phase transition. In fact, the close relation between 
the density of the hadronic medium, a percolation threshold, and the
occurrence of the QCD phase transition finds support in the numerical
simulations of QCD with varying number of partonic degrees of freedom.
With increasing number of light (nearly massless) quark flavours the
phase transition has been found to occur at lower temperatures.
This seems to be natural for a phase transition controlled be a 
percolation threshold: QCD with $n_f$
massless quarks describes a world with $(n_f^2-1)$ Goldstone
particles (pions). At a given temperature the density of the hadronic
medium thus will be proportional to $(n_f^2-1)$ and a critical density at
which these hadrons {\it start overlapping} occurs earlier with larger
$n_f$. While the change in the critical temperature is relatively small
when changing the number of flavours from, in general, two to four, there 
is a large change between $n_f=0$ (pure gauge theory) and $n_f=2$.
The transition temperature in QCD with two light
quarks is found to occur at $T_c(n_f=2) \simeq 150$MeV,
while in the purely gluonic theory, $n_f \equiv 0$, it is found to be
substantially higher, $T_c(n_f=0) \simeq 230$MeV. This may also easily be
understood in terms of a percolation threshold: In
the absence of quarks there are no light hadrons, the lightest states are
glueballs with a mass of $O(1~{\rm GeV})$. One thus needs a rather large
temperature to excite enough of these heavy glueball states to reach a
critical density.

Besides this change in the
temperature scale and details of the phase transition itself -- the
transition is first order for $n_f=0$ and likely to be second order for
$n_f=2$ -- the temperature dependence of bulk thermodynamic quantities is
very similar. In Fig.1 we show recent results for the energy density,
$\epsilon$, and the pressure, $P$, in the $SU(3)$ gauge theory$^5$ ($n_f=0$)
as well as two-flavour QCD$^7$.
As can be seen, in both cases a substantial change in the effective
number of degrees of freedom occurs only in the vicinity of $T_c$.

\midinsert
\vskip 9.5 truecm
\vbox{\baselineskip=12pt \eightrm {\eightbf Figure 1.}
Energy density and pressure for the pure {\eightit SU}(3) gauge theory (a) and
two-flavour QCD (b) as a function of {\eightit T/T}$_c$. The pure gauge theory
results have been obtained from simulations on a large 32$^3
\times$6
lattice$^6$ while those for two-flavour QCD so far exist only for a
rather small lattice$^7$ (8$^3\times$4). For
more details on the
inherent finite size effects in these lattice calculations we refer to
the above papers as well as one of the reviews on finite temperature
lattice QCD$^5$.
}
\endinsert

The numerical results for the energy density suggest that up to
temperatures of about $0.9T_c$ the hadronic
medium is well described by a gas of hadrons, in which only
the lightest states contribute.
In the case of a $SU(2)$ gauge theory the critical energy density at
$T_c$ has been studied in great detail$^8$. Here it is found that the
deconfinement transition occurs at $\epsilon_c/T_c^4 \simeq 0.25$, which
is only a factor five larger than the energy density of a glueball gas.
Already below $0.9T_c$ higher excited states play no role and one
is left with a rather dilute glueball. A rough estimate yields values of
about (0.1-0.2) for the number of glueballs per hadronic volume.

The slow variation of the energy density and pressure below $0.9T_c$
shows that also the gluon condensate will vary only little with
temperature. The gluon condensate is directly related to the difference
$\epsilon -3P$,
$$
\langle G^2 \rangle_T = \langle G^2 \rangle_0 - (\epsilon -3P)~~,
\eqno{(1)}
$$
where $\langle G^2 \rangle_0$ denotes the gluon condensate at $T=0$.
The small variation with $T$, which can be deduced from Fig.1, is in
accordance with chiral perturbation theory, which suggests that the
temperature dependence of $\langle G^2 \rangle_T$ starts only at $O(T^8)$
(Ref.~3).

For the chiral condensate $\langle \bar\psi \psi \rangle$ chiral
perturbation theory suggests a stronger dependence on temperature,
A
$$
\langle \bar\psi \psi \rangle(T) = \langle \bar\psi \psi \rangle (0) \biggl[
1 - {n_f^2 - 1 \over n_f} \biggl({T^2 \over 12 f_\pi^2}\biggr) -
{n_f^2 - 1 \over 2 n_f^2} \biggl({T^2 \over 12 f_\pi^2}\biggr)^2 +O(T^6)
\biggr]~~.
\eqno{(2)}
$$

\midinsert
\vskip 9.5 truecm
\vbox{\baselineskip=12pt \eightrm {\eightbf Figure 2.}
The chiral condensate at finite temperature extrapolated to the chiral
limit ({\eightit m}$_q\equiv$0) and normalized to the
corresponding zero temperature value for various values of the
number of flavours (Fig.2a). In Fig.2b we show results from a simulation in
quenched QCD ({\eightit n}$_f$=0) for various values of the
quark mass. The
simulations have been performed at fixed value of the gauge coupling$^9$,
6{\eightit /g}$^2$=6.0, on a 32$^3\times$8
lattice, corresponding to {\eightit T}$\simeq$0.92{\eightit T}$_c$ and
on various low temperature lattices with sizes ranging from
24$^3\times$32 to 32$^3\times$64 (Ref.10).
The condensates $\langle \bar\psi \psi \rangle_1$ and
$\langle \bar\psi \psi \rangle_2$ are defined in Section 4.
No temperature dependence is visible.
}
\endinsert

In Fig.2 we show
a collection of results obtained for QCD with various number of flavours.
They suggest at most a weak dependence of the chiral condensate on
temperature. However, in particular in the case of QCD simulations with
light quarks this behaviour has to be confirmed in simulations on larger
lattices. In the case of pure $SU(3)$ gauge theory simulations on large
lattices have been performed$^9$ and confirm that there is no significant
temperature dependence of the chiral condensate up to $T \simeq 0.92 T_c$.
It thus seems to be conceivable that significant changes in the QCD
condensates occur only quite close to $T_c$.

\vglue 12pt
\line{\tenbf 3. Hadronic Properties close to $T_c$ \hfil}
\vglue 5pt
Information about hadron masses and decay constants can be extracted in
lattice simulations from the long-distance behaviour of correlation
functions of hadron operators 
$$
G_H(x) = \langle H(x) H^+ (0) \rangle \rightarrow e^{-m_H |x|}
\quad,\quad x\equiv (\tau, \vec x)~~,
\eqno{(3)}
$$
with $H(x)$ denoting an operator with the appropriate quantum numbers of
the hadronic state under consideration, for instance $H(x) = \bar\psi_u(x)
\gamma_5 \psi_d(x)$ for the pion. At zero temperature one studies
the behaviour of the correlation function at large Euclidean times $\tau$.
From the exponential decay of $G_H$ one deduces the hadron mass of the
lightest state in this channel, whereas the amplitude is related to the
decay constant of this hadronic state. At finite temperature the Euclidean
extent is limited, $0\le \tau \le 1/T$, and one thus studies the behaviour
of $G_H$ for large spatial separations, $|\vec x | \rightarrow \infty$.
This yields information about the finite temperature screening masses
which are related to pole masses as long as there is a bound state
in the quantum number channel under consideration.

Finite temperature screening masses have been studied in lattice
simulations of QCD for quite some time$^{11}$. A very drastic qualitative
change in the screening masses is seen when one crosses the QCD
transition temperature. Parity partners become degenerate above $T_c$,
the pseudo-scalar mass becomes massive and approaches the free
quark/anti-quark value, $m_{\rm meson} = 2\pi T$, at large temperatures.
These features do not seem to depend much on the number of quark
flavours. In particular, they also have been found in quenched QCD
simulations. It thus seems to be meaningful to first study the behaviour
of hadronic properties in the quenched approximation where results on
large lattices with high accuracy can be obtained. In the following
we will describe the results of such an investigation performed on a
rather large lattice ($32^3\times 8$) close to the phase transition
temperature.
\vglue 12pt
\line{\tenit 3.1. The GMOR Relation and the Pion Decay Constant \hfil}
\vglue 5pt
The GMOR relation relates the chiral condensate to the pion mass and
pion decay constant,
$$
f_\pi^2 m_\pi^2 = m_q \langle \bar\psi \psi \rangle_{m_q=0}~~.
\eqno{(4)}
$$
Below $T_c$ the pion is a Goldstone particle, its
mass squared depends linearly on the quark mass,
$$
m_\pi^2 = a_\pi m_q ~~.
\eqno{(5)}
$$
A calculation of the chiral condensate and the pion mass at different
values of the quark mass allows the determination of the pion slope,
$a_\pi$, and the zero quark mass limit of the condensate,
$\langle \bar\psi \psi \rangle_{m_q=0}$.

The pion decay constant $f_\pi$ can be determined directly from the
relevant matrix element
$$
\sqrt{2} f_\pi m_\pi^2 = \langle 0| \bar\psi_u \gamma_5 \psi_d |\pi^+
\rangle ~~.
\eqno{(6)}
$$

\midinsert
\vskip 9.5 truecm
\vbox{\baselineskip=12pt \eightrm {\eightbf Figure 3.}
Test of the GMOR relation at {\eightit T}=0.92{\eightit T}$_c$ in quenched 
QCD (a) and
the temperature dependence of the pion decay constant (b). In Fig.3a we
compare the results obtained from the amplitude of the pion correlation
function through Eq.(6) (circles) and extrapolated to zero quark mass with
the result obtained from the GMOR relation (star). Also shown is the zero
temperature result at this value of the gauge coupling (square).
}
\endinsert

The square of the matrix element appearing on the right hand side of
Eq.(6) is proportional to the amplitude of the pion correlation function
and can be determined in a Monte Carlo calculation.
A comparison of $f_\pi$ determined this way with the ratio
$\langle \bar\psi \psi \rangle_{m_q=0} / a_\pi$ thus provides a direct
test of the GMOR relation at finite temperature. This is shown in Fig.3a
at $T\simeq 0.92T_c$. Moreover, we can compare the
value of $f_\pi$ calculated at finite temperature with corresponding
zero temperature results. This is shown in Fig.3b. Clearly we do not have
any evidence for violations of the GMOR relation nor for a significant
change of $f_\pi$ with temperature below $T_c$. The sudden change above
$T_c$ reflects the drastic change in the structure of the pseudo-scalar
correlation function. It does no longer have a pole corresponding to a
Goldstone-particle (pion). The existence of a pseudo-scalar bound state above
$T_c$ thus is questionable. Certainly for large temperatures such a state
does not exist, the correlation function is dominated by a quark/anti-quark
cut. The notion of $f_\pi$ used for the square root of the amplitude of
the correlation function should thus be used with caution above $T_c$.

\vglue 12pt
\line{\tenit 3.2. Vector Meson Mass and Nucleon Mass \hfil}
\vglue 5pt
The possibility of a variation of the $\rho$ meson mass with temperature
has been discussed a lot as this might lead to modifications of
dilepton spectra$^1$, which are experimentally detectable.
The temperature dependence of the meson masses has been discussed within
the framework of the OPE. Arguments have been given that the meson masses
are temperature independent up to $O(T^2)$ (Ref.~4).
The Monte Carlo calculations of the vector meson correlation function at
finite temperature also show no significant temperature dependence  of the
mass even close to $T_c$. In Fig.4 we show the result of our
calculation of the screening mass in the vector channel correlation
function at $T\simeq 0.92T_c$ and compare this with zero temperature
calculations at the same value of the gauge coupling. There is no
evidence for any temperature dependence. The same holds true for the
nucleon mass, although the details are more subtle in this case. As can be
seen in Fig.4b there is a clear difference between the local masses,
$m_N(z) \sim \ln G_N(z)/G_N(z+1)$,
extracted on a large zero temperature lattice and those extracted at
finite temperature on a lattice of size $32^3\times 8$. However, as the
nucleon is a fermion, there is a non-negligible contribution from the
non-zero Matsubara mode to the nucleon screening mass,
$$
m_N =\sqrt{\tilde m_N^2 +(\pi T)^2}
\eqno{(7)}
$$
After removing the thermal contribution, $\pi T$, the nucleon mass,
$\tilde m_N$, agrees with the zero temperature result within
statistical errors.

\midinsert
\vskip 9.5 truecm
\vbox{\baselineskip=12pt \eightrm {\eightbf Figure 4.}
The $\rho$-meson mass in quenched QCD for various values of the quark mass
(a) and local masses from the nucleon correlation function (b). The
simulations have been performed at fixed value of the gauge coupling$^9$,
6{\eightit /g}$^2$=6.0, on a 32$^3\times$8
lattice corresponding to {\eightit T}$\simeq$0.92{\eightit T}$_c$ and
on various low temperature lattices (see Fig.2). 
In Fig.4b we show estimates for
the nucleon mass obtained from the nucleon correlation function at distance
{\eightit z}. These local masses converge to the nucleon mass in the
limit {\eightit z}$\rightarrow \infty$. The horizontal band indicates the
corresponding extrapolation at zero temperature.
No temperature dependence is visible in the $\rho$-meson mass as well as
the nucleon mass (see text for appropriate subtraction of the finite
temperature Matsubara mode).
}
\endinsert

\vglue 12pt
\line{\tenbf 4. Two-Flavour QCD \hfil}
\vglue 5pt
The discussion of hadron masses and decay constants in quenched QCD shows
that there are no visible deviations from the zero temperature values up
to temperatures $T\simeq 0.9T_c$. One may expect that this is partly due to
the fact that the phase transition in pure $SU(3)$ gauge theory is first order.
The change from the hadron phase to the quark gluon plasma phase is
more abrupt than in the case of a second order transition and might lead
to less significant changes in condensates and hadronic properties.

The presently available simulations of QCD with two light quark flavours,
indeed, suggest that the transition is smoother. In fact, the
analysis of the scaling behaviour of various thermodynamic quantities
with the quark mass seems to be consistent with the behaviour expected for
a second order chiral phase transition$^{12}$. We thus may wonder
whether stronger modifications of hadron properties can be expected in this
case.

\midinsert
\vskip 9.5 truecm
\vbox{\baselineskip=12pt \eightrm {\eightbf Figure 5.}
The chiral condensate (a) and the chiral susceptibility (b) in two flavour
QCD obtained from simulations on an 8$^3\times$4
lattice$^{12}$.
}
\endinsert

In Fig.5 we show the behaviour of the chiral condensate and its derivative
with respect to the quark mass -- the chiral susceptibility $\chi_m$,
$$\eqalign{
\langle \bar\psi \psi \rangle_1 &= {n_f \over 4} {T \over V} 
{\partial \over \partial m_q} \ln Z \cr
\chi_m &= {T \over V} {\partial^2 \over \partial m_q^2} \ln Z \cr}~~.
\eqno{(8)}
$$
Also shown there is an improved estimator for the zero quark mass chiral
condensate in which the $O(m_q)$ contribution to $\langle \bar\psi \psi
\rangle_1$ is removed,
$$
\langle \bar\psi \psi \rangle_2 =\langle \bar\psi \psi \rangle_1 -
m_q \chi_m~~.
\eqno{(9)}
$$
As can be seen $\langle \bar\psi \psi \rangle_1$ deviates from $\langle
\bar\psi \psi \rangle_2$ strongly only in a narrow region around $T_c$.
This indicates that only in this region contributions from the singular
part of the free energy, which is responsible for the occurrence of a phase
transition, is dominant. Also the chiral susceptibility,
$\chi_m$, has a very narrow peak at the pseudo-critical point.
We thus expect that also in the case of two-flavour QCD substantial
changes in hadronic properties will occur only in a narrow temperature
regime close to $T_c(m_q)\equiv T_{\rm peak}$,
$$
\Delta_{\rm crit} = \biggl| {T_{\rm peak/2}-T_{\rm peak} \over T_{\rm peak}}
\biggr| \lsim 0.1~~.
\eqno{(10)}
$$

\vglue 12pt
\line{\tenbf 5. Conclusions \hfil}
\vglue 5pt
We have discussed some of the basic properties of QCD thermodynamics,
which are relevant for a discussion of the temperature dependence of
hadronic properties in the hadronic phase of QCD. There are strong
indications that the gluon as well as the chiral condensates show at most
a weak dependence on temperature for $T\lsim 0.9T_c$. So far no
statistically significant temperature dependence could be observed in
lattice QCD calculations. The same is true for hadron masses and decay
constants calculated in quenched QCD.

At least for the chiral condensate one expects, however, a strong
variation with temperature for $0.9T_c < T < T_c$. This is supported by
the currently available lattice calculation in two-flavour QCD. The
temperature variation of the gluon condensate, on the other hand, might
not be that strong as indicated by the behaviour of the equation of state
shown in Fig.1. The influence of this on the behaviour of hadronic
parameters will certainly be investigated in the near future.

\vglue 12pt
\line{\tenbf Acknowledgements \hfil}
\vglue 5pt
The numerical work that has been reviewed here has been performed on the
Cray YMP at the HLRZ-J\"ulich and the QUADRICS parallel
computer at the University of Bielefeld, funded by DFG under contract No.
Pe 340/6-1. I thank all my collaborators in
these projects (Refs. 6 and 9) for many discussions and their support in
the preparation of this review. In particular, I thank J. Engels,
G. Boyd and E. Laermann for making the numerical results of these projects
presentable.
\vglue 12pt
\line{\tenbf References \hfil}
\vglue 5pt
\baselineskip=11pt
\frenchspacing
\item{1.} F.~Karsch, K.~Redlich and L.~Turko,
{\it Z. Phys. C - Particles and Fields} {\bf 60} (1993) 519.
\item{2.} for a review and further references see for instance: M.A.
Shifman in {\it QCD 20 years later, Vol. 2}, ed. P.M. Zerwas and H.A. Kastrup
(World Scientific, Singapore 1993).
\item{3.} for a review and further references see for instance: H.
Leutwyler in {\it QCD 20 years later, Vol. 2}, ed. P.M. Zerwas and H.A. Kastrup
(World Scientific, Singapore 1993).
\item{4.} V.L. Eletsky and B.L. Ioffe, {\it Phys. Rev.} {\bf D47} (1993)
3083;
\item{  } C.A. Dominguez and M. Loewe, {\it Comment on Current Correlators
in QCD at Finite Temperature}, UCT-TP-208-94
and references therein.
\item{5.} For a review and further references see for instance: F. Karsch,
{\it Nucl. Phys.} {\bf B (Proc. Suppl.) 34} (1994) 63 and F. Karsch in
{\it QCD 20 years later, Vol. 2}, ed. P.M. Zerwas and H.A. Kastrup
(World Scientific, Singapore 1993).
\item{6.} G. Boyd, J. Engels, F. Karsch, E. Laermann, C. Legeland,
M. L\"utgemeier and B. Petersson, {\it in preparation}.
\item{7.} T. Blum, L. K\"arkk\"ainen, D. Toussaint and S. Gottlieb, 
{\it The Equation of State for Two-Flavour QCD}, contribution to the {\it
XII International Symposium on Lattice Field Theory (LATTICE 94)}, to 
appear in {\it Nucl. Phys. B (Proc. Suppl.)}. 
\item{8.} J. Engels, F. Karsch and K. Redlich,
{\it Scaling Properties of the Energy Density in $SU(2)$ Lattice
Gauge Theory}, BI-TP 94/30.
\item{9.} G. Boyd, S. Gupta, F.~Karsch, E.~Laermann, B.~Petersson
and K.~Redlich, {\it Hadron Properties just before Deconfinement},
BI-TP 94/42.
\item{10.}  see Ref.6 for detailed references.
\item{11.} C. DeTar and J.B. Kogut, {\it Phys. Rev.} {\bf D36} (1987) 2828;
\item{  } K.D. Born et al., {\it Phys. Rev. Lett.} {\bf 67} (1991) 302.
\item{12.} F. Karsch and E. Laermann,
{\it Susceptibilities, the Specific Heat and a Cumulant in Two-Flavour QCD},
BI-TP 94/29, to appear in {\it Phys. Rev. D}.

\vfil\supereject
\bye